# A new material for hydrogen storage; ScAl$_{0.8}$Mg$_{0.2}$

Martin Sahlberg[a], Premysl Beran[b], Thomas Kollin Nielsen[c], Yngve Cerenius[d], Krisztina Kádas[e,f], Marko P. J. Punkkinen[g], Levente Vitos[e,f,g], Olle Eriksson[e], Torben R. Jensen[c], Yvonne Andersson[a]

[a]*Department of Materials Chemistry, Uppsala, University, Uppsala, Sweden*
[b]*Nuclear Physics Institute ASCR v.v.i. and Research Centre Rez Ltd., 25068 Rez, Czech Republic*
[c]*Interdisciplinary Nanoscience Center (iNANO) and Department of Chemistry, University of Århus, Aarhus C, Denmark*
[d]*MAX-lab, Lund University, S-22100 Lund, Sweden*
[e]*Department of Physics and Materials Science, Uppsala University, Box 530, 75121 Uppsala, Sweden*
[f]*Research Institute for Solid State Physics and Optics, P.O. Box 49, H-1525 Budapest, Hungary*
[g]*Department of Materials Science and Engineering, Royal Institute of Technology, SE-10044 Stockholm, Sweden*

**Abstract**

A novel aluminium rich alloy for hydrogen storage has been discovered, ScAl$_{0.8}$Mg$_{0.2}$, which has very promising properties regarding hydrogen storage capacity, kinetics and stability towards air oxidation in comparison to hydrogen absorption in state-of-the-art intermetallic compounds. The absorption of hydrogen was found to be very fast, even without adding any catalyst, and reversible. The discovered alloy crystallizes in a CsCl-type structure, but decomposes to ScH$_2$ and Al(Mg) during hydrogen absorption. Detailed analysis of the hydrogen absorption in ScAl$_{0.8}$Mg$_{0.2}$ has been performed using *in situ* synchrotron radiation powder X-ray diffraction, neutron powder diffraction and quantum mechanical calculations. The results from theory and experiments are in good agreement with each other.

**Introduction**
Light metal hydrides are the only hydrogen storage materials with a potential to meet the numerous demands for an energy carrier in vehicles. Magnesium and aluminium based materials have been extensively studied within the past 40 years [1-3]. Magnesium has been explored intensively due to the high theoretical gravimetric and volumetric hydrogen density, virtually limitless amounts on the Earth and in addition, magnesium is a cheap metal. Unfortunately its utilization is hampered by unfavourable thermodynamics. Therefore, a high number of Mg-based alloys have been investigated for improved properties [4].

The hydrogen absorption in $Mg_{0.65}Sc_{0.35}$ has been studied by Latroche *et al.* [5, 6]. They have shown that this compound absorbs deuterium (a favourable isotope for studies of hydrogen in solids) while transforming from a random CsCl-type structure to a $CaF_2$-type structure, leading to a final composition of 2.2 D/M. This compound was used with a small addition of palladium to ensure good kinetic properties of deuterium loading/unloading.

The solid solubility of magnesium in the Sc-Al compounds were previously studied by Gröbner *et al.* who showed that up to 20% aluminium could be replaced by magnesium in the ScAl-phase [7]. Furthermore, the hydrogen resistance of Al-Sc compounds was partially studied by Antonova & Chernogorenko [8], but no structural determinations or other types of characterizations of the hydride phases have been published to our knowledge. Also, hydrogen absorption in $Sc_2Al$ has been studied by Burkhanov et al. [9], where they showed that this compound absorbs hydrogen by decomposing to $ScH_2$ and Al. This reaction was found to be irreversible at the investigated temperatures and pressures.

Prior to this study, Sc-Al-Mg alloys have not been considered for hydrogen storage. In this paper we present the first study of the structural transformation during hydrogen absorption of $ScAl_{0.8}Mg_{0.2}$. The reaction mechanism between the alloy and hydrogen gas has been studied using in situ synchrotron radiation X-ray powder diffraction (SR-XRD) as well as first principles theory based on density functional theory.

**Materials and methods**

*Experimental*
$ScAl_{0.8}Mg_{0.2}$ was prepared by heating appropriate amounts of the elements, Mg (99.95 %, Alfa Aesar), Sc (99.99%, Rare Earth products Ltd) and Al (99.999 %, Gränges SM). The elements were placed inside a tantalum crucible which was sealed under argon atmosphere. The tube was heated in an induction furnace up to ~1000 °C in argon atmosphere ($p$(Ar) = 40 kPa). The tube was opened in a glovebox to avoid oxygen contamination. No reaction between the sample and the Ta crucible was observed.

The grounded sample was hydrogenated by solid-gas reactions at different hydrogen pressures and temperatures, 2-10 MPa and 20-400 °C. A short exposure to air during loading of the reactor was unavoidable but the reactor was evacuated and flushed with argon several times prior to hydrogenation to avoid oxide formation at elevated temperatures. No activation of the material prior to hydrogen absorption was needed.

Powder X-ray diffraction data for structural analysis was recorded on a X-ray powder diffractometer (XRD), Bruker D8 powder diffractometer equipped with an Våntec PSD (4 degree opening) using Cu Kα$_1$ radiation. Silicon was used as internal calibration standard. Crystal structure refinements were performed using the Rietveld method [10] implemented in the program Fullprof [11].

*In situ* SR-PXD measurements were performed at the MAX-lab Synchrotron in Lund, Sweden, using the Beamline I711 [12]. The wavelength used was refined to 1.09994 Å and the X-ray exposure time was 30 s/scan. The diffracted intensities were measured using a Mar165 CCD detector. A sapphire single crystal tube was used as a sample holder and a gas supply system allowing changes in both gas and pressure via a vacuum pump was attached to the sample cell [13]. The sample cell was heated by resistive heating (tungsten wire) and the temperature was measured by a thermocouple placed inside the sapphire tube. This setup allows changes of pressure and temperature during the measurements [14].

The crystal structure of the deuterated sample was determined at room temperature using neutron powder diffraction (NPD) pattern recorded at LWR15 reactor in Rez near Prague. The sample was contained in a vanadium cylinder. Three bended Si(422) single crystals were used as monochromator with wavelength of 1.274 Å. Diffractograms between 5-120° in 2Θ were collected with step of 0.1°. Refinements were performed on a deuterated sample of ScAl$_{0.8}$Mg$_{0.2}$ ($p$(D$_2$) = 4 MPa and $T$ = 400 °C).

*Theory*
The ab initio calculations are based on the density functional theory [15] formulated within the generalized gradient approximation (GGA) [16] for the exchange-correlation functional applying the Exact Muffin-tin Orbitals (EMTO) method [17,18]. The substitutional disorder of the metal and H atoms was taken into account using the Coherent Potential Approximation (CPA) [19,20] as implemented in the EMTO-CPA method [21]. In the self-consistent calculations, the one-electron equations were treated within the scalar relativistic and soft core approximations. The EMTO Green's function was calculated for 32 energy points. In the EMTO basis set *s*, *p*, *d*, *f* and *g* orbitals were included. In the irreducible part of the Brillouin zones 165 *k*-points were used in the case of the CsCl structure and 89 *k*-points in the case of the CaF$_2$ and the quasi-random structures. The total charge density was expanded in spherical harmonics, including terms up to $l_{max}$=8.

Additional test calculations for pure Sc and ScH$_2$ (structure STR-IV) were carried out using the full-potential Vienna ab initio simulation package (VASP) [22-25] in combination with the projector augmented wave method [26,27] and GGA [16]. The two mixing enthalpies for ScH$_2$ obtained with VASP (-0.146 Ry) and EMTO (-0.141 Ry) are found to be very close to each other, which confirms that EMTO is a suitable tool to study the energetics of the ScAlMg-H system. Since the EMTO method was also used to perform the calculations for the random systems, here we present only results obtained using this approach.

**Results**
*Experiment*
ScAl$_{0.8}$Mg$_{0.2}$ crystallizes with cubic symmetry and the diffraction pattern shown in Fig. 1 has been indexed with the unit cell parameter a=3.4106(1) Å. The crystal structure was determined to a pseudo binary CsCl-type structure, Pm-3m, where Sc occupied the 1a site (0, 0, 0) and Al and Mg the 1b site (0.5, 0.5, 0.5). However, the structure refinements showed mixed occupancy on the 1a site, 90% Sc and 10% Al/Mg and on the 1b site, 90% Al/Mg and 10% Sc.

ScAl$_{0.8}$Mg$_{0.2}$ was hydrogenated *in situ* at a hydrogen pressure of 10 MPa and a temperature increase from 20 to 400 °C at a heating rate of 10 °C /min. The temperature was kept at 400 °C for the rest of the experiment. According to Fig. 2 the hydrogen absorption started just before 400 °C and the alloy decomposes to ScH$_2$ and Al(Mg).

After hydrogen absorption the hydrogen pressure was released to vacuum, which is indicated in the upper part in Fig. 2. The hydrogenated phase was stable at 400 °C in vacuum. However, from *ex situ* experiments it is evident that all hydrogen was released for temperatures above 480 °C in vacuum and the intermetallic CsCl-type structure was reformed, see below.

In Fig. 3 the transition from CsCl-type structure to ScH$_2$ and Al(Mg) is shown. The different phases co-exist until the whole material has turned into ScH$_2$ and Al(Mg). This is in agreement with our calculations (see below) and with earlier studies on similar types of compounds [9, 28]. For the currently investigated system the transition from the CsCl-type structure to ScH$_2$ and Al(Mg) is rapid. The total time for the phase transformation was ~3 min.

Refinement of the neutron diffraction pattern was used to confirm the crystal structure of the deuterated sample, as shown in Fig. 4. This analysis showed that the hydrogenated phase was indeed ScH$_2$ which crystallizes in space group Fm-3m (CaF$_2$ structure), where the Sc atom sits at the 4a position (0, 0, 0) and deuterium on the 8c site (0.25, 0.25, 0.25), a = 4.7774(1) Å. The 8c position was completely filled with deuterium, leading to the final composition ScD$_2$. This result was in agreement with gravimetric measurements.

The hydrogen absorption in ScAl$_{0.8}$Mg$_{0.2}$ is completely reversible, as shown in Fig. 5. When heating the sample under vacuum all hydrogen is desorbed for temperatures ~500 °C. This is a very interesting result as pure ScH$_2$ does not desorb hydrogen below 960 °C [29]. The ScH$_2$ is significantly destabilized by the formation of ScAl$_{0.8}$Mg$_{0.2}$. These results are also in good agreement with our calculations, discussed below.

*Calculation.*

First we investigate the theoretical Gibbs energy of formation

$$\Delta G'(T,x) = G(T,x) - 0.5xG_2(T) - (1-0.5x)G'_0(T) \qquad (1)$$

for the random alloy (Sc$_{0.5}$Al$_{0.4}$Mg$_{0.1}$)H$_x$ in the CaF$_2$ structure (Fig. 6) calculated as a function of hydrogen content (*x*) for temperatures T=0 K (solid circles) and 700 K (open circles). In Eqn. (1), we chose as standard states the random Sc$_{0.5}$Al$_{0.4}$Mg$_{0.1}$ alloy in the FCC structure with Gibbs energy $G'_0(T) = E'_0 - TS'_0$ and the random (Sc$_{0.5}$Al$_{0.4}$Mg$_{0.1}$)H$_2$ alloy in the CaF$_2$ structure with Gibbs energy $G_2(T) = E_2 - TS_2$. The total energies $E'_0$ and $E_2$ and the corresponding equilibrium lattice parameters are listed in Table 1. In Eqn.1 G(T,x) is the Gibbs energy of (Sc$_{0.5}$Al$_{0.4}$Mg$_{0.1}$)H$_x$. Here and in the following we assume that for the solids in question the main temperature effect in the Gibbs energy is represented by the configuration entropy $S$. Since these solids possess similar Debye temperatures, the error in $\Delta G'$ introduced by the above approximation is estimated to be negligible.

The negative curvature of $\Delta G'(0\text{K}, x)$ (black dots in Fig. 6) indicates the instability of the intermediate random compositions against segregation into H-poor and H-rich systems. Intermediate compositions might only appear as a result of entropy or short range order effects. We find that the Gibbs energy of formation $\Delta G'(T, x)$ becomes negative around 500 K, and shows a small positive curvature at T=700 K with $\Delta G'(700\text{K}, 1) \approx -1.7 \text{mRy}$ (open dots in Fig.6).

In order to assess the effect of clustering or ordering in the $CaF_2$ lattice, we investigated the Gibbs energies of four quasi-random structures calculated for $x$=0, 1 and 2. In the first case (denoted by STR-I), Sc and $Al_{0.8}Mg_{0.2}$ occupy the $1a$ and $1d$ sites, respectively, of the $L1_0$ lattice and the hydrogen atoms are randomly distributed in the tetrahedral positions with mixed coordination. The STR-I structure models the possible ordering of Sc and Al-Mg sites. The hydrogen-induced segregation is modelled by using a supercell formed by two adjacent FCC unit cells, the first filled up with Sc and the second with $Al_{0.8}Mg_{0.2}$. Then for $x$=1 the originally random distribution of hydrogen atoms (50% site occupancy) is modified so that the hydrogen atoms are moved from the tetrahedral positions coordinated by $Al_{0.8}Mg_{0.2}$ to the tetrahedral positions coordinated by Sc (structure STR-II). We also considered the opposite, hydrogen atoms on the tetrahedral positions around AlMg (structure STR-III). Finally, a possible phase-separation for $x$=1 is modelled by a composite formed by grains of $ScH_2$ ($CaF_2$) and $Al_{0.8}Mg_{0.2}$ (random FCC) (structure STR-IV). The Gibbs formation energies for the above four cases are denoted by $\Delta G'(T,$ STR-I), $\Delta G'(T,$ STR-II), $\Delta G'(T,$ STR-III) and $\Delta G'(T,$ STR-IV) in Fig. 6. Note that for $x$=0 and 2, STR-II and STR-III are equivalent structures. We would like to point out that the decomposed system seen in experiment should be somewhere between the STR-II and STR-IV model structures. However, since both the energetics associated with the grain boundaries and the diffusion barriers are completely neglected in STR-IV, we expect that the STR-II is the best approximant for the real situation.

For $x$=0, the STR-I structure is found to be ~9-12 mRy (depending on temperature) more stable than the disordered FCC structure. However, $\Delta G'(T,$ STR-I) is gradually reduced with hydrogen addition, indicating that the $L1_0$ ordering tendency is diminished by the presence of H. In contrast to STR-I, the STR-II structure is obtained to be unstable for $x$=0. On the other hand, amongst the structures considered, this phase turns out to be favourable for $x$=1 and 2. In particular, for 50% hydrogen content, $\Delta G'(T,$ STR-II$) \approx -20 \text{mRy}$. Comparing this energy to +20 mRy obtained for the STR-III structure at $x$=1, we conclude that there is a very strong hydrogen-induced Sc segregation at intermediate compositions. Namely, the hydrogen atoms prefer the Sc-coordination rather than the $Al_{0.8}Mg_{0.2}$-coordination. This is also confirmed by the ideal phase-separated system (STR-IV, where all hydrogen is located in Sc), which yields $\Delta G'(0$ K, STR-IV$) \approx$ -33 mRy. As will become clear later, the above short-range order effect is in fact the main atomic-level driving force for the observed hydrogenation of $ScAl_{0.8}Mg_{0.2}$.

Next we study the reaction between $ScAl_{0.8}Mg_{0.2}$ in the CsCl structure and $H_2$, *viz.*

$$0.5 \ ScAl_{0.8}Mg_{0.2} + 0.5x H_2 \rightarrow (Sc_{0.5}Al_{0.4}Mg_{0.1})H_x. \qquad (2)$$

The Gibbs energy of the reaction can be cast into the form

$$\Delta G(T,x) = \Delta G'(T,x) - \Delta G''(T,x), \qquad (3)$$

where, depending on the structure of the hydrogenated sample, $\Delta G'$ is either the Gibbs energy of the random $(Sc_{0.5}Al_{0.4}Mg_{0.1})H_x$ given in (1) or the Gibbs energy of one of the four quasi-random structures considered above. In (3) we have introduced the notation

$$\Delta G''(T,x) \equiv \frac{1}{2} x \left[ G_{H_2}(T) - G_2(T) + G'_0(T) \right] + \left[ G_0(T) - G'_0(T) \right], \qquad (4)$$

where $G_{H_2}(T)$ is the Gibbs energy of hydrogen reservoir [30],

$$G_{H_2}(T) = E_{H_2} + k_B T \left[ \ln\left( \frac{p_{H_2}}{k_B T n_Q} \right) - \ln Z_{int} \right]. \qquad (5)$$

Here $E(H_2)$ is the experimental energy of a $H_2$ molecule, $p_{H_2}$ is the pressure of the $H_2$ gas, the quantum concentration is $n_Q = (mk_B T / 2\pi \hbar^2)^{3/2}$ ($m$ stands for the mass of a $H_2$ molecule, $\hbar$ is the Planck constant, $k_B$ is the Boltzmann constant), and $Z_{int}$ is the partition function of the internal states due to the rotational and vibrational degrees of freedom. The first term in the right hand side of Eqn. (4) represents the binding energy difference for two H atoms in the $H_2$ molecule and in random $(Sc_{0.5}Al_{0.4}Mg_{0.1})H_2$, and the second term is the difference between the Gibbs energy of the disordered $Sc_{0.5}Al_{0.4}Mg_{0.1}$ in the FCC structure ($G'_0$) and that of partially ordered $ScAl_{0.8}Mg_{0.2}$ in the CsCl structure ($G_0 = E_0 - TS_0$). The calculated $E_0$ and the corresponding lattice parameter are given in Table 1.

With the above notations, reaction (2) becomes exothermic if

$$\Delta G'(T,x) \leq \Delta G''(T,x). \qquad (6)$$

The two relative Gibbs energies are compared in Fig. 6 for $T \leq 700$ K and $p_{H_2}$ = 10 MPa and 100 kPa, corresponding approximately to the experimental conditions. We find that at 0 K the hydrogenated random alloy is stable for $x \geq 1.8$ and the STR-I, STR-II and STR-IV structures are stable for $x=1$ and 2. Nevertheless, due to the large hydrogen entropy (~0.1 Ry/K per $H_2$ molecule), $\Delta G''(T,x)$ exhibits a much stronger temperature dependence than $\Delta G'(T,x)$, which results in a significant temperature effect on the hydrogen uptake Gibbs energy $\Delta G(T,x)$.

During hydrogenation, we assume that the hydrogen gas has constant pressure $p_{H_2}$ = 10 MPa. According to Fig. 6, at this pressure the reaction for $x=2$ is exothermic for temperatures below ~300 K, where both the STR-I and STR-II structures can form. No H-rich compositions appear above 300 K. However, for $x=1$ the stability field of hydrogenated alloys is significantly larger. We find that the STR-II structure for $x=1$ remains stable for temperatures up to ~550 K, above which the reaction becomes endothermic. On the other hand, the ideal STR-IV composite is predicted to remain stable for temperatures up to ~800 K.

During the hydrogen release, the pressure of the hydrogen gas is below the atmospheric pressure. Using the upper limit $p_{H_2}$ = 100 kPa, we find that the transition temperatures for reaction (1) are shifted to ≤200 K at $x$=2 and to ~400 K at $x$=1. Therefore, according to our calculations based on the STR-II model structure, the $(Sc_{0.5}Al_{0.4}Mg_{0.1})H_x$ alloys with $x \approx 1$ are stable at ambient conditions and a temperature increase of about 100 °C is needed to start the hydrogen release process. We note that for the ideal STR-IV case, the hydrogen release starts at significantly larger (~600 K) temperature.

**Discussion**

We show here for the first time that $ScAl_{0.8}Mg_{0.2}$ absorbs hydrogen while decomposing to $ScH_2$ and Al(Mg). This reaction occurs at a hydrogen pressure of 10 MPa and a temperature close to 400 °C. These results show the possibility to find new light metal alloys with a reversible hydrogen storage capacity of more than 2 wt%. $ScAl_{0.8}Mg_{0.2}$ showed no significant degradation after storing in air for three months. In contrast to the previously reported results on $Sc_2Al$ [9] and electrochemically deuterated $Mg_{0.65}Sc_{0.35}$ [5] the hydrogen absorption in $ScAl_{0.8}Mg_{0.2}$ is fully reversible and also very rapid. These important properties are obtained without adding a catalyst. Furthermore, hydrogen absorption occurred in a one step reaction.

The $ScAl_{0.8}Mg_{0.2}$ alloy is very stable and easy and safe to handle in air which is of great importance for practical applications. We expect that further research will lead to modifications of the presented alloy that release hydrogen at lower temperatures. $ScAl_{0.8}Mg_{0.2}$ has interesting properties regarding hydrogen storage capacity and kinetics in comparison to hydrogen absorption in state-of-the-art intermetallic compounds [1, 4]. Therefore, there is a good perspective for future applications with this material, and it may provide inspiration for design of other novel classes of materials.

We also found that results from experiments and calculations are in good agreements with each other in pointing out the advantage of $ScAl_{0.8}Mg_{0.2}$ as a hydrogen storage material. $ScAl_{0.8}Mg_{0.2}$ exhibits the CsCl type structure, with only a small mixing of the elements. This was shown to be the most stable configuration, by comparing with both random CsCl (BCC) and FCC. It is also interesting to see that for the ideal STR-IV case, the experimental hydrogen release temperature is close to the calculated value. The somewhat higher temperature observed in experiments could be due to kinetic barriers or to the approximations employed in the theoretical study and it is the topic of future studies.

**Acknowledgements**
We recognize the support from The Swedish Research Council and The Nordic Energy Programme. M.S. acknowledges financial support from the Royal Swedish Academy of Sciences and Helge Ax:son Johnsons stiftelse. K.K. and L.V. acknowledge financial support from the Hungarian Scientific Research Fund (T048827, K-68312), and P.B. acknowledges financial support from Research Centre Rez (AV0Z10480505, MSM2672244501).

**Figure captions**

Fig.1: (Color online.) Rietveld refinement of the $ScAl_{0.8}Mg_{0.2}$ structure. Si is used as internal calibration standard, lower Bragg positions ($\lambda$=1.540598 Å ).

Fig. 2: Hydrogenation of $ScAl_{0.8}Mg_{0.2}$ investigated by *in situ* SR-PXD. The structure transformation occurs around 400 °C. The pressure is shown on the right side of the figure.

Fig. 3: *In situ* SR-PXD data that show the decomposition from CsCl-type structure to $ScH_2$ and Al(Mg).

Fig. 4: (Color online.) Rietveld refinement of the sample deutered at 4 MPa and 400 °C. The upper and lower tics marks the Bragg positions of $ScD_2$ and Al(Mg) respectively.

Fig. 5: (Color online.) XRD pattern of the as synthesized, hydrogenated and desorbed sample, showing the reversibility. Si is used as internal calibration standard.

Fig. 6: (Color online.) Comparison between the Gibbs energies (per formula unit) $\Delta G'(T,x)$ [Eq. (1)], $\Delta G'(T,\text{STR-I})$, $\Delta G'(T,\text{STR-II})$, $\Delta G'(T,\text{STR-III})$ and $\Delta G'(T,\text{STR-IV})$ (see text for definitions), and $\Delta G''(T,x)$ [Eq. (4)] calculated for the $ScAl_{0.8}Mg_{0.2}$-H system for $T \leq 700$ K and $p_{H_2} = 10$ MPa and 100 kPa.

Table 1. Theoretical equilibrium lattice parameters ($a_0$) and total energies ($E$) of $Sc_{0.5}Al_{0.4}Mg_{0.1}$ in the FCC structure ($E_0'$), $Sc_{0.5}Al_{0.4}Mg_{0.1}H_2$ in $CaF_2$ structure ($E_2$), and $ScAl_{0.8}Mg_{0.2}$ in the CsCl structure ($E_0$). All energies are per formula unit.

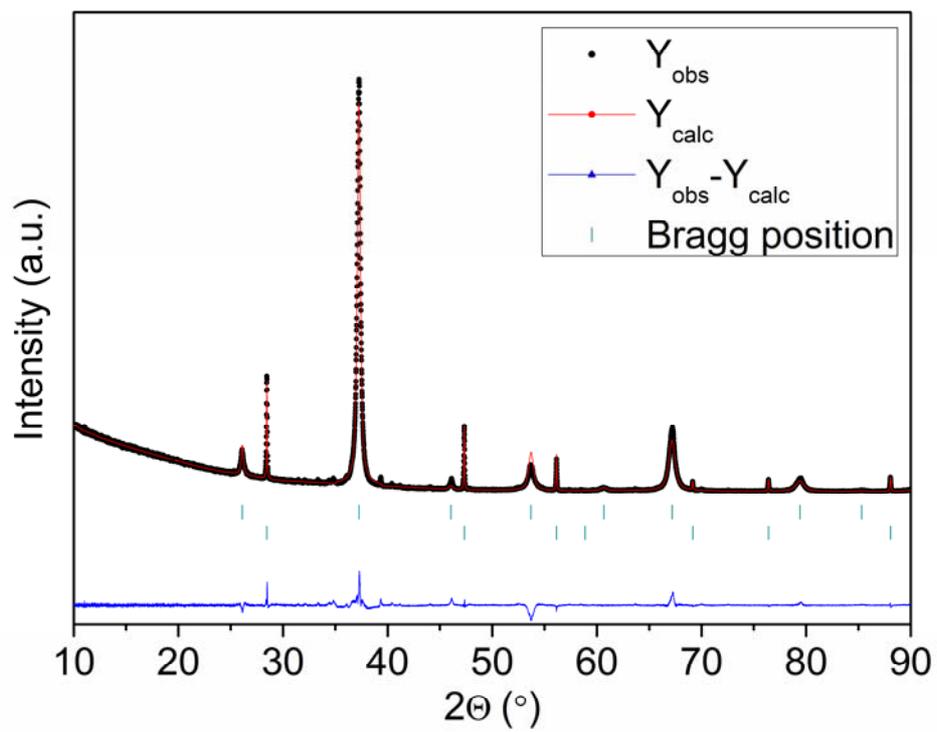

Figure 1

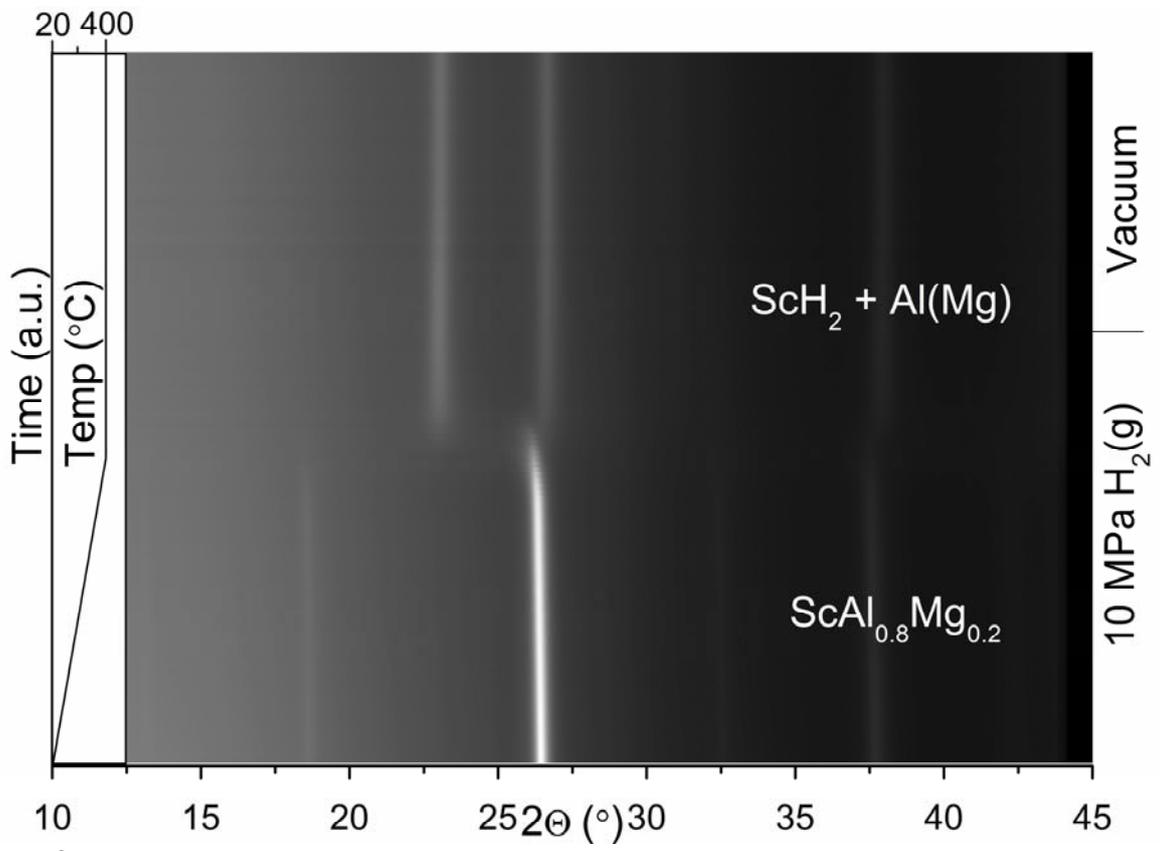

Figure 2

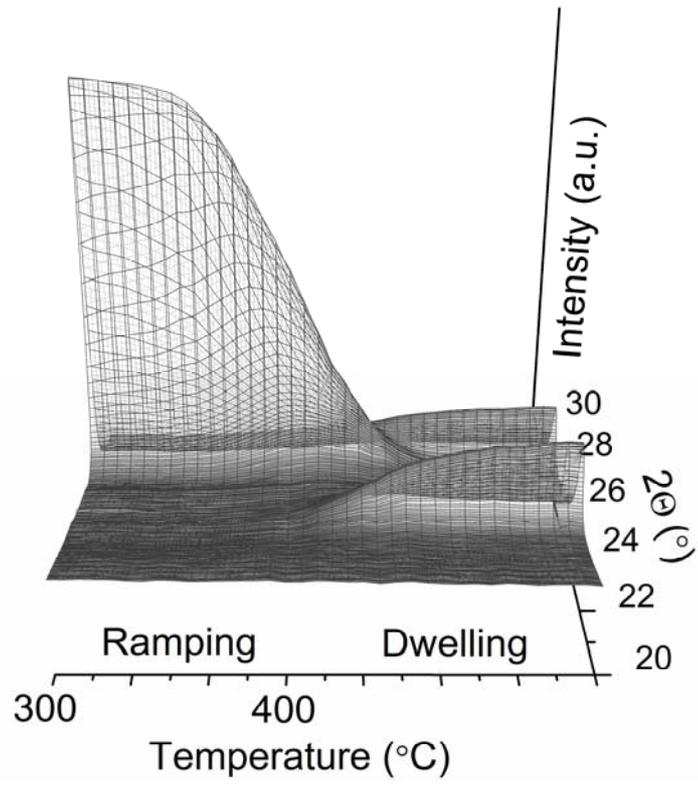

Figure 3

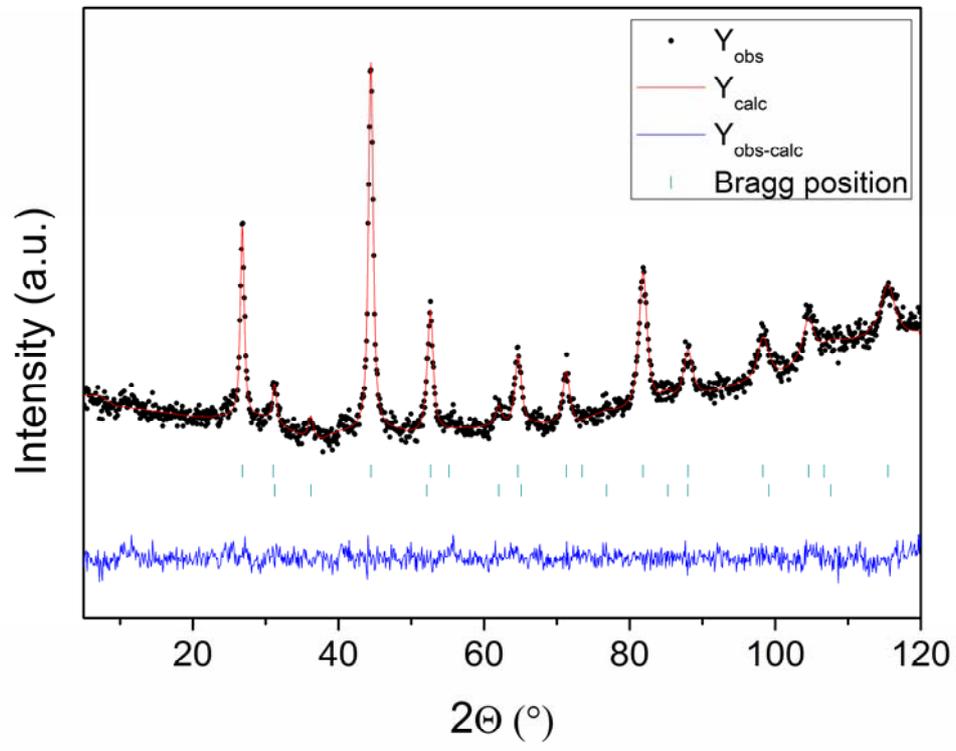

Figure 4

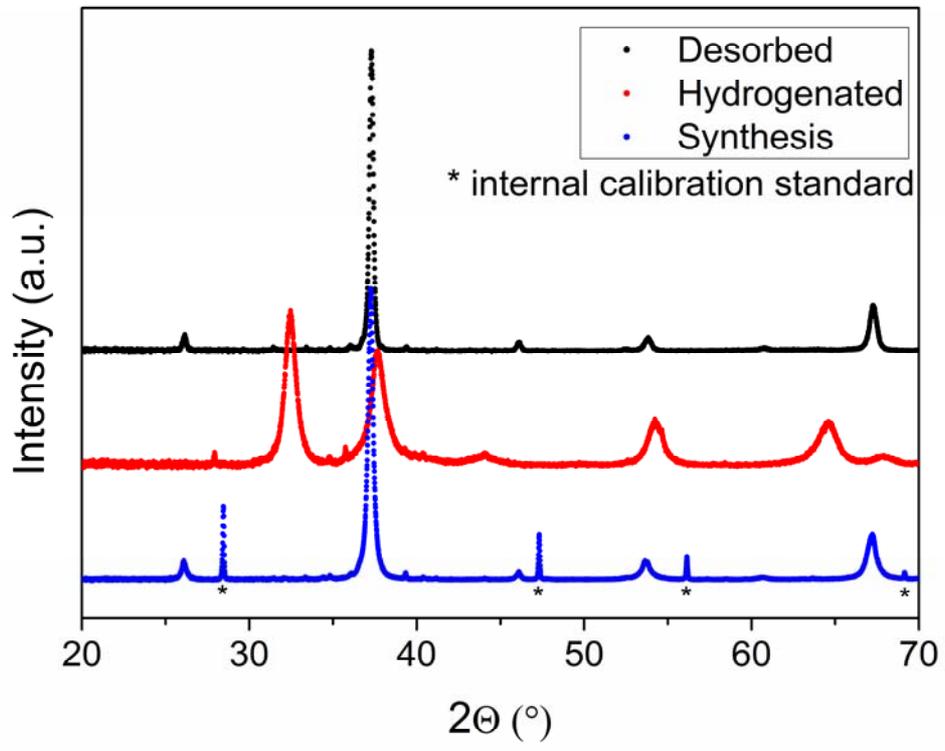

Figure 5

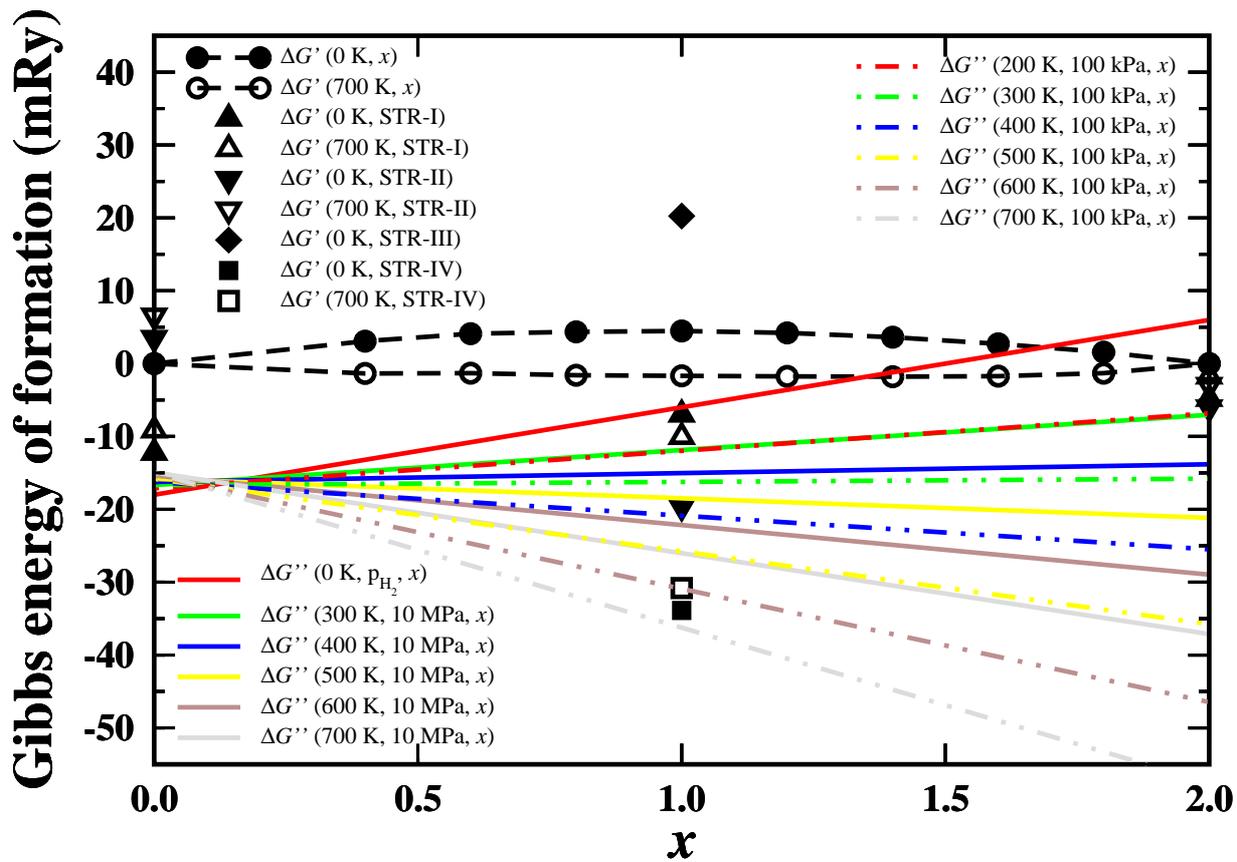

Figure 6

|                             | structure | $a_0$ (Å) | energy (Ry)        |
|-----------------------------|-----------|-----------|--------------------|
| $Sc_{0.5}Al_{0.4}Mg_{0.1}$      | FCC       | 4.363     | $E'_0=-998.557$    |
| $Sc_{0.5}Al_{0.4}Mg_{0.1}H_2$   | $CaF_2$   | 4.746     | $E_2=-1000.906$    |
| $ScAl_{0.8}Mg_{0.2}$            | CsCl      | 3.421     | $E_0=-998.568$     |